\documentclass[published]{JHEP3} 

\JHEP{00(2010)000}

\usepackage{color}
\let\normalcolor\relax

\usepackage{ulem}
\usepackage{epsfig,multicol,bbm,amsmath}

\newcommand{\lsim}{\mathrel{\mathop{\kern 0pt \rlap
  {\raise.2ex\hbox{$<$}}}
  \lower.9ex\hbox{\kern-.190em $\sim$}}}
\newcommand{\gsim}{\mathrel{\mathop{\kern 0pt \rlap
  {\raise.2ex\hbox{$>$}}}
  \lower.9ex\hbox{\kern-.190em $\sim$}}}

\newcommand\fverb{\setbox\fverbbox=\hbox\bgroup\verb}
\newcommand\fverbdo{\egroup\medskip\noindent%
			\fbox{\unhbox\fverbbox}\ }
\newcommand\fverbit{\egroup\item[\fbox{\unhbox\fverbbox}]}
\newbox\fverbbox

\newcommand{\bea}{\begin{eqnarray}}
\newcommand{\eea}{\end{eqnarray}}

\newcommand{\svds}{\langle \sigma v \rangle_{\rm ds}}
\newcommand{\svfo}{\langle \sigma v \rangle_{\rm fo}}

\title{Dark Matter that can form Dark Stars}

\author{Paolo Gondolo\\
	\\ Department of Physics and Astronomy, University of Utah
	\\ 115 South 1400 East, Salt Lake City, UT 84112, USA
	Email: \email{paolo.gondolo@utah.edu}}
\author{Ji-Haeng Huh \\
	Department of Physics and Astronomy, Seoul National University\\
	Seoul, Korea, 151-747\\
	E-mail: \email{jhhuh@phya.snu.ac.kr}}
\author{Hyung Do Kim \\
	FPRD and Department of Physics and Astronomy, Seoul National University\\
	Seoul, Korea, 151-747\\
	E-mail: \email{hdkim@phya.snu.ac.kr}}
\author{ Stefano Scopel\\
	Department of Physics, Sogang University\\
        Seoul, Korea, 121-742\\
	E-mail: \email{scopel@sogang.ac.kr}}

\received{\today} 		
\accepted{\today}		

\abstract{The first stars to form in the Universe may be powered by
  the annihilation of weakly interacting dark matter particles. These
  so-called dark stars, if observed, may give us a clue about the
  nature of dark matter. Here we examine which models for particle
  dark matter satisfy the conditions for the formation of dark
  stars. We find that in general models with thermal dark matter lead
  to the formation of dark stars, with few notable exceptions: heavy
  neutralinos in the presence of coannihilations, annihilations that
  are resonant at dark matter freeze-out but not in dark stars, some
  models of neutrinophilic dark matter annihilating into neutrinos
  only and lighter than about 50 GeV. In particular, we find that a
  thermal DM candidate in standard Cosmology always forms a dark star
  as long as its mass is heavier than $\simeq$ 50 GeV and the thermal
  average of its annihilation cross section is the same at the
  decoupling temperature and during the dark star formation, as for
  instance in the case of an annihilation cross section with a non--vanishing
  $s$-wave contribution.}

\keywords{Dark Star, Thermal Dark Matter, MSSM, leptophilic, neutrinophilic}

\begin{document}


\section{Introduction}

The first stars, also referred to as Population III stars, are the
first luminous objects in the Universe. They contribute to the
reionization of the interstellar medium, they provide the heavy
elements (metals) that eventually become part of the later generations
of stars, and they may be the seeds of the very massive black holes
observed in quasars.

It was shown in \cite{dark_stars_prl,Freese:2008hb,Spolyar:2009nt} that the first stars to form in
the Universe may be powered by the annihilation of dark matter
particles instead of nuclear fusion. These dark-matter powered stars,
or dark stars for short, constitute a new phase of stellar
evolution. Besides the assumption that dark matter is made of weakly
interacting massive particles (WIMPs) that can self-annihilate into
ordinary particles, three conditions are necessary for the formation
of a dark star.

The first condition is that the density of dark matter at the location of the (proto)star must be high enough for dark matter to efficiently and rapidly annihilate into ordinary
particles, releasing a large amount of energy. The first stars are believed to form at the center of dark matter halos when the Universe was young (redshift $z\sim 10$-50) and denser than today. Not only the dark matter density at the center of those early halos was high, but as the baryonic gas contracted into the first protostars, more dark matter was gathered around the forming object by the deepening of the gravitational potential (gravitational contraction). Cosmological parameters and the evolution of the gas density completely determine the resulting density of dark matter at the location of the first protostars. Analytic and numerical evaluations \cite{dark_stars_prl,Freese:2008hb,Natarajan:2008db} lead to a resulting density which is high enough to satisfy the first condition for the formation of a dark star.

The second condition is that a large fraction of the energy released
in the dark matter annihilation must be absorbed in the gas that
constitutes the (proto)star. The fraction $f_Q$ of annihilation energy
deposited into the gas depends on the nature of the annihilation
products. Typical products of WIMP annihilation are charged leptons,
neutrinos, hadrons, photons, W and/or Z bosons, and Higgs bosons. The
latter (W, Z, and Higgs) decay rapidly into leptons and hadrons. The
hadrons themselves, which are mostly charged and neutral pions) decay
rapidly into charged leptons, neutrinos, and photons (although a small
number of stable particles like protons can also be produced). After
$\sim 10^{-8}$ seconds, all unstable elementary particles, including
the muon, have decayed away, and only protons, electrons, photons and
neutrinos survive. Protons have a large scattering cross section with
the protostar medium and are quickly absorbed. Electrons and photons
can ionize the medium and/or generate electromagnetic showers. For
WIMPs with mass $m \gtrsim 0.5$ GeV, electromagnetic showers are the
dominant process. At the time when the following third condition for a
dark star is satisfied, the protostar has a diameter of more than 40
radiation lengths, implying that all the energy released in protons,
electrons, and photons is absorbed inside the protostar. Only the
fraction of energy carried away by the neutrinos is lost for what
concerns a dark star.

The third condition for the formation of a dark star is that the
heating of the (proto)star gas arising from the dark matter
annihilation energy must dominate over any cooling mechanism that
affects the evolution of the (proto)star. In \cite{dark_stars_prl}, it
was shown that the dark matter heating rate $Q_{\rm DM}$, in energy
deposited per unit time and unit volume, is given by the expression
\begin{equation}
Q_{\rm DM} = f_Q \frac{\svds}{m} \rho^2 ,
\end{equation}
where $\rho$ is the dark matter density inside the (proto)star, which
is determined by the cosmological model, and $\svds$ is the average
value of the dark matter annihilation cross section $\sigma$ times
WIMP relative velocity $v$ inside a dark star. To the extent that
electromagnetic showers are generated, i.e.\ $m \gtrsim 0.5$ GeV, all
dark star properties depend on the particle physics model only through
the quantity
\begin{equation}
f_Q \frac{\svds}{m}.
\end{equation}

Ref.~\cite{dark_stars_prl} fixed the annihilation cross section to $\svds=3\times10^{-26}$ cm$^3$/s and examined a range of WIMP masses $m$ from 1 GeV to 10 TeV.  In addition, Ref.~\cite{dark_stars_prl} assumed $f_Q = 2/3$,
based on simulations of neutralino dark matter annihilation in the
Minimal Supersymmetric Standard Model (MSSM). For this range of $Q_{\rm DM}$, they compared the heating and cooling rates along protostar evolution tracks from \cite{yoshida}, and concluded that there is a time during the evolution of the protostar in which the dark matter heating dominates over all cooling rates. This finding lead to the realization that dark stars may be possible.

In this paper, we examine the possible values of $Q_{\rm DM}$ for a
large selection of particle physics models, and verify if the third
condition above is satisfied in these models. We find that not all
particle dark matter models lead to the formation of dark stars,
although the models that do not form dark stars are either tuned to
resonant annihilation or rather artificial.

\begin{figure}
\begin{center}
\includegraphics[width=0.7\linewidth]{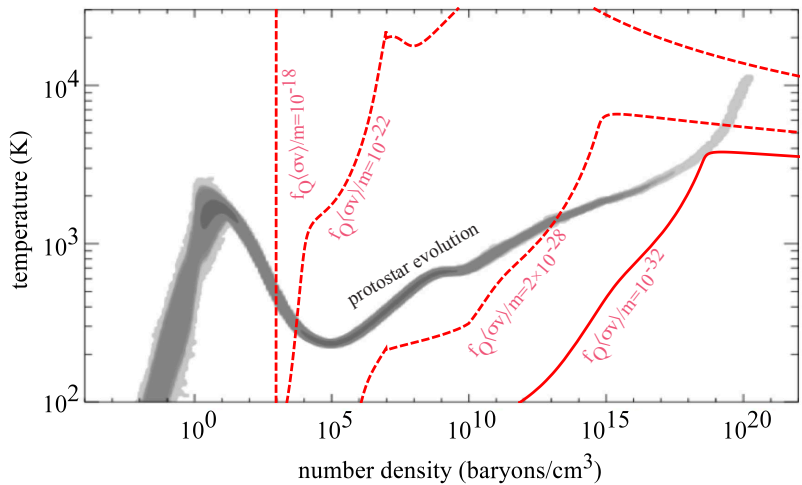}
\end{center}
\caption{Condition for the formation of a dark star, in terms of the protostar gas density and temperature. The gray band shows possible evolution tracks of the protostar obtained through numerical simulations in a $\Lambda$CDM cosmology \cite{yoshida}. The red lines show critical curves on which the heating rate from dark matter annihilation equals the total cooling rate of the protostar gas. Critical curves are labeled by the value of $f_Q \svds/m$ in units of cm$^3$ s$^{-1}$ GeV$^{-1}$. A dark star forms at the intersection of a critical line with the gas evolution track. No dark star can form for $f_Q \svds/m < 10^{-32}$ cm$^3$ s$^{-1}$ GeV$^{-1}$ (solid critical line on the right).}
\label{fig:thermo}
\end{figure}

The restriction imposed by the third condition for dark star formation is best expressed in terms of a condition on the quantity $f_Q\langle \sigma v\rangle_{\rm ds}/m$. Following \cite{dark_stars_prl}, we have computed the critical lines in the gas temperature-density plane at which the heating rate from dark matter annihilation equals the total cooling rate. These lines are shown in Figure \ref{fig:thermo} for a wide range of values of $f_Q\langle \sigma v\rangle_{\rm ds}/m$, from $10^{-18}$ cm$^3$ s$^{-1}$ GeV$^{-1}$ to $10^{-32}$ cm$^3$ s$^{-1}$ GeV$^{-1}$. Below the latter value, the heating-cooling critical line no longer intersects the thermodynamic track of the protostellar gas, indicated by the gray band obtained through numerical simulations of the formation of the first stars in a $\Lambda$CDM cosmology \cite{yoshida}. In other words, for $f_Q\langle \sigma v\rangle_{\rm ds}/m < 10^{-32}$ cm$^3$ s$^{-1}$ GeV$^{-1}$, the protostar is expected to contract to a regular Population III star powered by nuclear fusion without passing through the dark star phase. At the other side of the $f_Q\langle \sigma v\rangle_{\rm ds}/m$ range, the critical line reaches a limiting curve given by the vertical line labeled $f_Q\langle \sigma v\rangle_{\rm ds}/m = 10^{-18}$ cm$^3$ s$^{-1}$ GeV$^{-1}$. Larger values of $f_Q\langle \sigma v\rangle_{\rm ds}/m$ give the same vertical line. Thus, as expected, if the annihilation rate is large, a protostar passes through the dark star phase. Therefore the third condition for the formation of a dark star is
\begin{equation}
f_Q \frac{\langle \sigma v\rangle_{\rm ds}}{m} > 1\times 10^{-32} \hbox{ cm$^3$ s$^{-1}$ GeV$^{-1}$}.
\label{eq:ds_condition}
\end{equation}

The choice $\svds=3\times10^{-26}$ cm$^3$/s in \cite{dark_stars_prl}
was motivated by the assumption that the dark matter WIMPs are
produced thermally in the early Universe. That is, that the WIMPs are
generated in matter-antimatter collisions at temperatures higher than
$T_{\rm fo} \sim m/20$, which is the temperature after which WIMP
production ``freezes out'' and the comoving WIMP number density
remains (approximately) constant. Ref.~\cite{dark_stars_prl} used the
following simple inverse-proportionality relation between the present
WIMP density $\Omega_\chi$ and the annihilation cross section $\svfo$
at the time of WIMP freeze-out,
\begin{equation}
\Omega_\chi h^2 = \frac{3\times 10^{-27} {\rm cm^3/s}}{\svfo} .
\label{eq:Oh2}
\end{equation}
Furthermore, ref.~\cite{dark_stars_prl} simply assumed that the
velocity-averaged annihilation cross section times relative velocities
at the time of freeze-out and in a dark star have the same value,
$\svds=\svfo$. In reality, the relation between $\Omega_\chi$ and
$\svfo$ is more complex, and in addition $\svds$ may differ from
$\svfo$ because $\sigma v$ may depend sensitively on the WIMP velocity
$v$. In this regard, we notice that the average WIMP speed at
freeze-out is of the order of
\begin{equation}
v_{\rm fo} \sim \sqrt{\frac{T_{\rm fo}}{m}}\sim \frac{c}{\sqrt{20}} \sim7\times10^{4}{\rm~km/s},
\end{equation}
while the typical speed of WIMPs in a dark star can be estimated from the orbital velocity
\begin{equation}
v_{\rm ds} \sim \sqrt{\frac{GM}{r}} \sim {\rm 30~to~300~km/s},
\end{equation}
namely $\sim 30$ km/s for a newly-born 1-$M_\odot$ dark star of 1 AU radius or $\sim 300$ km/s for a mature 600-$M_\odot$ dark star of 5 AU radius.

A neutralino in the MSSM provides an example of a more complex
relation between $\Omega_\chi$ and $\svfo$. At the same time, it
allows the direct evaluation of both $\svds$ and $f_Q$, and in general
it has $\svds\ne\svfo$. Section~\ref{sec:mssm} explores this case.

Kaluza-Klein dark matter is examined in Section 3, where it is concluded that generically in these models $\svds$ tends to be larger or comparable to $\svfo$.

Leptophilic models of dark matter proposed to explain the PAMELA
positron excess and the Fermi and HESS cosmic-ray electron-positron
data provide another example in which $\svds$ may not be the same as
$\svfo$. They are examined in Section~\ref{sec:lepto}.

Finally, we push $f_Q$ down using dark matter particles that
annihilate exclusively into neutrinos (``neutrinophilic'' models). In
these models, even if annihilation produces predominantly neutrinos
that escape the forming star, W- and Z-bremsstrahlung processes may
generate enough charged leptons to actually form a dark star. We
examine this case in Section~\ref{sec:neutro}.

\section{MSSM}
\label{sec:mssm}

Because of its many free parameters (more than 100), the MSSM provides
a variety of examples in which the annihilation cross section in the
dark star differs from the annihilation cross section at the time of
freeze-out, or more precisely $\svds\ne\svfo$.

There are several ways in which the equality $\svds=\svfo$ can be
violated in the MSSM \cite{Griest:1990kh}. 
First, the quantity $\sigma v$ may depend on the
relative velocity $v$. This includes three cases: (i) $p$-wave
annihilation in which $\sigma v = a + b v^2$ is dominated by the $b
v^2$ term at freeze-out (here $a$ and $b$ are constants); (ii)
resonant annihilation in which $\sigma v$ follows a Breit-Wigner
function $(1+v^2)^{c}/[(v^2+\delta)^2+\gamma^2]$, where $\delta$,
$\gamma$ and $c$ are constants; and (iii) threshold annihilation in
which an annihilation channel is kinematically accessible at
freeze-out but not in a dark star thanks to the higher particle
kinetic energies at freeze-out. Second, the annihilation reactions
that determine the freeze-out time may be unrelated to the
neutralino-neutralino annihilation that occurs inside a dark star, in
that the freeze-out temperature may be high enough to convert
neutralinos into heavier supersymmetric particles that annihilate much
faster (a phenomenon called coannihilation). It is then the
annihilation cross section of the heavier supersymmetric particles
that determines the neutralino relic density, and this cross section
is in general not the same as the neutralino-neutralino annihilation
cross section, thus $\svds\ne\svfo$. We remark in passing that
resonant annihilation and coannihilations are not rare phenomena in
the MSSM, and are actually essential to obtain a neutralino dark
matter in the minimal supergravity or constrained MSSM models.

To illustrate these four cases ($p$-wave annihilation, resonant annihilation, threshold annihilation, and coannihilation), it is sufficient to consider a so-called effective MSSM (effMSSM) with eight free parameters fixed at the electroweak scale \cite{effMSSM}. These parameters are: the CP-odd Higgs boson mass $m_A$, the ratio of neutral Higgs vacuum expectation values $\tan\beta$, the Higgs mass parameter $\mu$, the gaugino mass parameters $M_1$ and $M_2$, the slepton mass parameter $m_{\tilde\ell}$, the squark mass parameter $m_{\tilde{q}}$, the ratios $A_{\tilde{\tau}}/m_{\tilde\ell}$, $A_{\tilde{t}}/m_{\tilde{q}}$ and $A_{\tilde{b}}/m_{\tilde{q}}$ involving the trilinear couplings  $A_{\tilde{\tau}}$, $A_{\tilde{t}}$ and $A_{\tilde{b}}$ of the third generation of sleptons and squarks (the three ratios are assumed to be equal).

We consider the parameter region of the effMSSM in which the lightest
neutralino is the lightest supersymmetric particle and its relic
density $\Omega_\chi$ is within the cosmological range $0.098 <
\Omega_\chi h^2 < 0.122$. In this region, we compute $\svds$ as the
value of $\sigma v$ at $v=0$. For each point in this region we also
compute $f_Q$ as the fraction of annihilation energy that does not go
into neutrinos. In obtaining $f_Q$, it is safe to assume that the
particle cascades after annihilation develop in vacuum, since muons,
taus and light mesons produced in the annihilation decay to neutrinos
before being stopped in the dark star medium.

\begin{figure}
\begin{center}
\includegraphics[bb=10 44 524 506,width=0.7\linewidth]{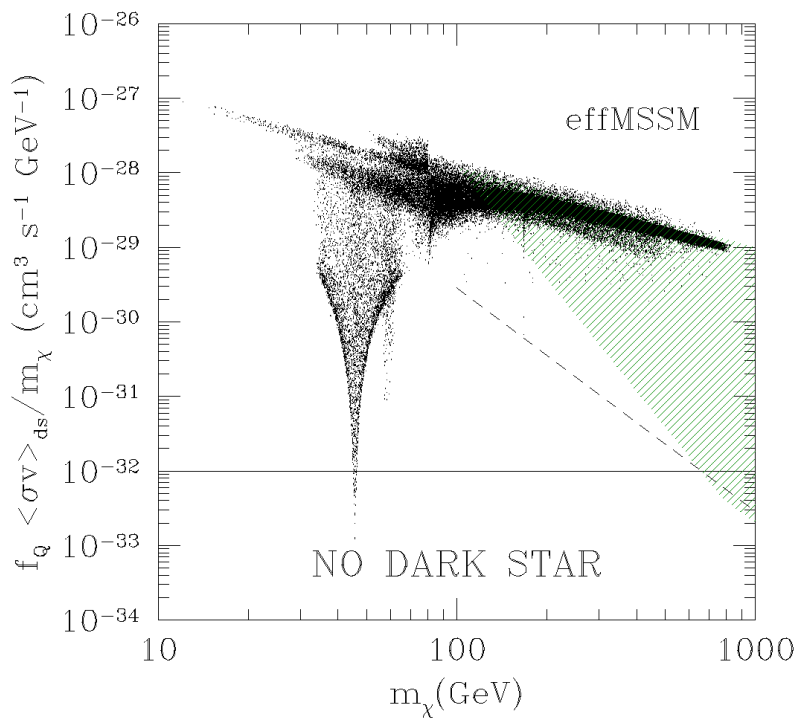}
\end{center}
\caption{Scatter plot of the combination $f_Q \svds/m_\chi$ plotted as
  a function of the neutralino mass $m_\chi$ for an effective
  eight-parameter MSSM model. Below the horizontal line a dark star
  cannot form. For resonant annihilation (the V-shaped feature at
  $m_\chi\sim 45$ GeV and the descending points around $m_\chi\sim60$
  GeV) and coannihilations (the shaded region and the dashed line on
  the right of $m_\chi \sim 100$ GeV), the quantity $f_Q \svds/m_\chi$
  may be too small for a dark star to form. For $p$-wave annihilation
  (the band sloping down to the right) and threshold annihilation (the
  various ``fingers'' of points dropping from the $p$-wave
  annihilation band), dark stars would be able to form.}
\label{fig:effmssm}
\end{figure}

Figure \ref{fig:effmssm} shows the values of the combination $f_Q
\svds/m_\chi$ obtained in the way just described as a function of the
neutralino mass $m_\chi$. There are four classes of points: (i) the
spread of points along the direction sloping down to the right is due
to $p$-wave annihilation; (ii) the V-shaped feature at $m_\chi \sim
45$ GeV is due to resonant annihilation through the Z boson (other
resonant annihilations, through the lightest Higgs boson of mass
varying from 115 GeV to 120 GeV, are visible at $m_\chi \sim 60$ GeV),
(iii) the ``fingers'' of points dropping from the $p$-wave band of
points arise from threshold coannihilation, and (iv) the shaded region
on the right of $m_\chi \sim 100$ GeV corresponds to possible
coannihilation with staus (the dashed line shows the similar boundary
for sneutrino coannihilations). These four cases are described in the
following.

In $p$-wave annihilation, the dominant contribution at freeze-out
comes from the $p$-wave term $b v^2$ in $\sigma v$. The $p$--wave
contribution to $\sigma v$, which is instrumental to provide the
correct neutralino relic density, is suppressed as far as the
evolution of the dark star is concerned. In fact, for a newly-born
1-$M_\odot$ dark star, one has $(v_{\rm ds}/v_{\rm fo})^2 \sim 2
\times 10^{-9}$. Thus in this case, $\svds \simeq a$ while $\svfo
\simeq b \langle v^2 \rangle\simeq b/20$, and in general they
differ. Their exact ratio depends on the particle physics parameters
contained in the coefficients $a$ and $b$. In our effMSSM scan,
$p$-wave annihilation gives rise to a spread in $f_Q\svds/m$ of about
one order of magnitude (band of points sloping down to the right in
Figure \ref{fig:effmssm}).

The Z resonance at $m_\chi \sim 45$ GeV provides an example of
resonant annihilation. The resonant part of the neutralino-neutralino
annihilation cross section is given by
\begin{equation}
(\sigma v)_Z = \beta_f\, \frac{g_{\rm eff}^4}{m_\chi^2}\,\frac{(s-m_Z^2)^2}{(s-m_Z^2)^2+\Gamma_Z^2m_Z^2} ,
\end{equation}
where $\beta_f$ is the speed of the final products in units of the
speed of light, and $g_{\rm eff}$ contains the coupling constants and
the mixing angles of the neutralinos and of the final particles
involved. The velocity dependence of $(\sigma v)_Z$ can be obtained by
writing $s=4m_\chi^2 (1+v^2)$, from which one finds, neglecting the
mass of the final products,
\begin{equation}
(\sigma v)_Z = \frac{g_{\rm eff}^4}{m_\chi^2}\,\frac{ (v^2+\delta)^2}{(v^2+\delta)^2+\gamma^2} ,
\label{eq:resonance}
\end{equation}
where $\delta=1-m_Z^2/(4m_\chi^2)$ and $\gamma=\Gamma_Z
m_Z/(4m_\chi^2)$. On resonance, that is for $2m_\chi=m_Z$ or
$\delta=0$ and $\gamma=\Gamma_Z/m_Z=0.0273$, the velocity-averaged
$\langle (\sigma v)_Z \rangle$ has very different values at freeze-out
($v\simeq0.2c$) and in a dark star ($v\simeq0$). At freeze-out, the
thermal average of $(\sigma v)_Z$ on resonance is, using $T_{\rm
  fo}=m_\chi/20$, $\langle (\sigma v)_Z \rangle_{\rm fo} =
0.97\,g_{\rm eff}^4/m_\chi^2$. On the other hand, in a dark star, one
has on resonance, for $v=30$ km/s, $\langle (\sigma v)_Z \rangle_{\rm
  ds} = 1.3\times10^{-13} g_{\rm eff}^4/m_\chi^2$. While $\langle
(\sigma v)_Z \rangle_{\rm fo} \sim 3\times10^{-26}$ cm$^3$/s to
provide the correct relic density, $\langle (\sigma v)_Z \rangle_{\rm
  ds}$ is thirteen orders of magnitude smaller. These very different
values of $\svfo$ and $\svds$ give rise to the V-shaped feature in
Figure~\ref{fig:effmssm} around $m_\chi=m_Z/2\sim 45$ GeV. (Similar
resonant features through the lightest Higgs boson appear superposed
at $m_\chi =m_H/2\sim 60$ GeV.)

Threshold annihilation occurs when the neutralino mass is slightly
smaller than half the total mass of the final annihilation products in
a specific channel (for example, $\chi\chi\to WW$). In this case,
kinetic energy is required for the reaction to occur. This kinetic
energy is available at the time of freeze-out thanks to the relatively
high temperature of the neutralinos, but is not available at the lower
velocities of neutralinos in a dark star. Therefore, the annihilation
into the specific channel ($\chi\chi\to WW$ in the example) occurs at
freeze-out but not in a dark star. The cross section $\svds$ is thus
smaller than $\svfo$. In Figure~\ref{fig:effmssm} this is illustrated
by the ``fingers'' of points dropping from the $p$-wave band at
$m_\chi \sim 80$ GeV (the $WW$ channel) and $m_\chi\sim190$ GeV (the
$t\bar{t}$ channel). In neither case the suppression of $\svds$ is
severe enough to bring the points outside the parameter region in
which dark stars can form.

For coannihilations, the relic density is determined by an effective
annihilation cross section $\langle \sigma v \rangle_{\rm eff}$, which
is an average of the annihilation cross sections of all reactions
between the neutralino and the coannihilating particles. In minimal
supergravity models, which are a subset of MSSM models,
coannihilations occur in specific regions of the parameter space in
which the stau $\tilde\tau$ is very close in mass to the neutralino
$\chi$, and with more tuning of the parameters when the stop
$\tilde{t}$ is very close in mass to $\chi$. In the general MSSM,
coannihilations may also occur between the lightest and second
lightest neutralino, and between the neutralino and the chargino.

For the sake of illustration in the context of dark stars, we focus on
stau coannihilations, because the experimental lower bound on the stau
mass ($m_{\tilde\tau} \gsim 98$ GeV) is smaller than the lower bound on
squark masses and thus the coannihilation region in parameter space is
larger. In the case of stau coannihilations, the effective
annihilation cross section is (approximately)
\begin{equation}
\langle \sigma v\rangle_{\rm eff} = \frac{ \langle \sigma v
  \rangle_{\chi\chi} + \langle \sigma v \rangle_{\chi\tilde\tau} \,
  e^{-(m_{\tilde\tau}-m_\chi)/T_{\rm fo}} + \langle \sigma v
  \rangle_{\tilde\tau\tilde\tau} \,
  e^{-2(m_{\tilde\tau}-m_\chi)/T_{\rm fo}}} {1 +
  e^{-(m_{\tilde\tau}-m_\chi)/T_{\rm fo}} +
  e^{-2(m_{\tilde\tau}-m_\chi)/T_{\rm fo}}} .
\label{eq:sigmaeff}
\end{equation}
Here $\langle \sigma v \rangle_{\chi\chi} $, $\langle \sigma v
\rangle_{\chi\tilde\tau} $ and $ \langle \sigma v
\rangle_{\tilde\tau\tilde\tau}$ are the total annihilation cross
sections for $\chi\chi\to{\rm anything}$, $\chi\tilde\tau\to{\rm
  anything}$, and $\tilde\tau\tilde\tau\to{\rm anything}$,
respectively. Since reactions like $\tilde\tau\tilde\tau\to\tau\tau$
are electromagnetic processes, $\langle \sigma v
\rangle_{\tilde\tau\tilde\tau} \sim \alpha^2/m_{\tilde\tau}^2 $, which
is much larger than the cross section for $\chi\chi\to\tau\tau$,
$\langle \sigma v \rangle_{\chi\chi} \sim \alpha^2
m_\tau^2/m_{\tilde\tau}^4$. In fact, for $m_{\tilde\tau}=100$ GeV (1
TeV), their ratio is approximately $\langle \sigma v
\rangle_{\tilde\tau\tilde\tau}/\langle \sigma v \rangle_{\chi\chi}
\sim m_{\tilde\tau}^2/m_\tau^2 \gtrsim 3\times10^3$
($3\times10^5$). Thus with an appropriate choice of the mass
difference $m_{\tilde\tau}-m_\chi$ in Eq.~(\ref{eq:sigmaeff}), one can
obtain an effective annihilation cross section three or more orders of
magnitude larger than the $\chi\chi$ annihilation cross section, and a
relic density three or more orders of magnitude smaller than without
coannihilations. This argument allows us to estimate a lower limit on
$\svds$ in a dark star using just the annihilation cross section for
$\chi\chi\to\tau\tau$ without having to compute the relic density in
the presence of coannihilations. The annihilation cross section for
$\chi\chi\to\tau\tau$ can be computed analytically and can be limited
from below by keeping only the diagram with $\tilde\tau$ exchange and
choosing appropriate neutralino and stau mixings. In this way, we
obtain
\begin{equation}
 \langle \sigma(\chi\chi\!\to\!\tau\tau) v \rangle_{\rm ds} \ge \frac{\pi\alpha^2}{32\cos^4\theta_W} \, \frac{m_\tau^2}{m_{\tilde\tau}^4}.
 \label{eq:sigmacoann}
\end{equation}
Then we set $m_{\tilde\tau}=m_\chi$ as appropriate for stau
coannihilations.  Moreover, bremsstrahlung
($\chi\chi\to\tau\tau\gamma$) gives a contribution to $\svds$ that
exceeds the lower limit just computed in Eq.~(\ref{eq:sigmacoann}) at
large neutralino masses. For $m_\chi=m_{\tilde\tau}$ we estimate
\begin{equation}
 \langle \sigma(\chi\chi\!\to\!\tau\tau\gamma) v \rangle_{\rm ds}
 \simeq \frac{\alpha^3}{m_\chi^2}.
 \label{eq:sigmacoann2}
\end{equation}
In addition, we compute $f_Q$ by examining the fraction of energy that
escapes into neutrinos in the decay chains of the $\tau$ lepton.
Eqs.~(\ref{eq:sigmacoann}) and~(\ref{eq:sigmacoann2}) are used to plot
the shaded region to the right of $m_\chi \sim 100$ GeV in
Figure~\ref{fig:effmssm}, namely
\begin{equation}
f_Q \frac{\svds}{m_\chi} \ge
\begin{cases}
\displaystyle 1.2\times10^{-28} \, \frac{\rm cm^3}{\rm s~GeV} \,
\left( \frac{\rm 100~GeV}{m_\chi} \right)^5, & \hbox{for
  $m_\chi\lesssim 800$ GeV} \\ & \\ \displaystyle 2\times10^{-30} \,
\frac{\rm cm^3}{\rm s~GeV} \, \left( \frac{\rm 100~GeV}{m_\chi}
\right)^3, & \hbox{for $m_\chi\gtrsim 800$ GeV} .
\end{cases}
\end{equation}
For the bremsstrahlung of gamma rays, we take $f_Q=1$. Dark stars can
form for $m_\chi \le 880$ GeV.  If $f_Q \le 0.86$, the bremsstrahlung
does not play any role in determining what is the largest possible
mass of neutralino forming dark stars, and dark stars can form for
$m_\chi \le 830$ GeV.  The annihilation cross section in dark stars
can be as low as the lower edge of this shaded region, while the
correct relic density is obtained through a much larger effective
annihilation cross section. We notice that in most of the shaded
region dark stars can still form, except at the higher masses where
the shaded region crosses the boundary of the area marked `no dark
star.'

Other coannihilations may arise in the MSSM. For instance, one might have coannihilations between the neutralino and a sneutrino or a selectron or a smuon. These may lead to even smaller $\svds$ than the case of stau coannihilations we use as an example, and so lead to a situation in which dark stars do not form. The worst case for dark stars is coannihilatin with sneutrinos, in that neutrinos are generated in the final state and neutrinos escape from the forming protostar without depositing energy. Similarly to the neutrinophilic case discussed below, three-body annihilation channels need to be considered, in particular internal and final state bremsstrahlung of charged leptons. A simple estimate of Z bremsstrahlung in the final state gives us

\begin{equation}
f_Q \frac{\langle \sigma(\chi\chi\to\nu\nu Z) v \rangle}{m_\chi}
\simeq 3 \times10^{-30} \, \frac{\rm cm^3}{\rm s~GeV} \, \left(
\frac{\rm 100~GeV}{m_\chi} \right)^3 .
\label{eq:sneutrino}
\end{equation}

Here we took $f_Q=1/2$ as a representative value. In the process of virtual internal bremsstrahung, Z can take away a sizable fraction of the energy and $f_Q$ may be sizable. Eq.~(\ref{eq:sneutrino}) is plotted in Figure~\ref{fig:effmssm} as the dashed line near the edge of the shaded coannihilation region. In terms of dark star formation, this case is similar to coannihilation with the $\tilde\tau$. Dark stars can form up to $m_\chi = 1$ TeV.
For different choice of $f_Q$, the cross point is at $m_\chi = (2f_Q)^{\frac{1}{3}}$ TeV ($m_\chi = 800$ GeV for $f_Q=1/4$.)

We therefore conclude that except in very special cases, namely on top
of the Z resonance or for coannihilations of heavy sleptons or
sneutrinos ($m_\chi\gtrsim 800$ GeV), dark stars can form in the MSSM.

\section{Kaluza-Klein Dark Matter}
\label{sec:kkdm}

If the Standard Model lives in five or six dimensions and the extra
dimensions are compactified at a radius $\simeq$ 1/TeV, the
Kaluza-Klein (KK) number can be preserved in a consistent way with all
the interactions involving an even number of odd KK number
particles. In this setup, the lightest KK particle (LKP) is stable and
can be a good dark matter candidate \cite{Servant:2002aq}.  In particular,
there are two interesting KK candidates for the dark matter: the KK photon (more
precisely the KK modes of the $U(1)_Y$ gauge boson) and the KK
neutrino.

The KK photon annihilates to quarks and leptons through the $t$-channel
exchange of KK fermions, and its relic abundance is compatible to
observation for its mass around 1 TeV.  If the right-handed KK electron,
muon, and tau are nearly degenerate (i.e if the mass difference is
$\lsim$1\%), the KK mass needed for the right relic density can drop to
700 GeV, since in this case the coannihilation cross section is very
small compared to the self annihilation cross section, leading to a
smaller effective cross section.

As far as the KK neutrino is concerned, its annihilation cross section
to quarks and leptons proceeds via $t$- or $s$-channel exchange of gauge
bosons, while annihilations to gauge bosons are mediated by $t$-channel
KK lepton exchange or $s$-channel gauge bosons. If one flavor of the KK
neutrino is considered, the correct relic density is obtained for a
mass around 1.5 TeV.  Including three flavors the effective cross
section becomes smaller due to coannihilations between different
flavors and the mass leading to the correct relic density is around 1
TeV.  An additional coannihilation process with the KK left-handed
electron is also possible when the latter has a smaller mass splitting
with the KK neutrino, but this effect is almost negligible.

In the case of KK dark matter the $s$-wave anniilation cross section is
always sizable both for the KK photon and for the KK neutrino, so that
there is little difference between $\langle \sigma v \rangle_{\rm ds}
$ and $\langle \sigma v \rangle_{\rm fo} $.  Moreover, the temperature
in the dark star is very low compared to the freeze out temperature,
so coannihilations with other particles give no contributions to the
effective cross section with the exception of exact degeneracy of the
masses. Anyway, since for KK dark matter the coannihilation cross
section is either smaller than the one without coannihilation or the
difference between the two is negligible, $\langle \sigma v
\rangle_{\rm ds} $ is expected to be always larger or comparable to
$\langle \sigma v \rangle_{\rm fo} $.  

Two interesting exceptions to the scenario described above are
resonant annihilation with level--2 KK particles \cite{Kakizaki:2005en}
and coannihilation with the KK gluon \cite{Burnell:2005hm}. In
principle these effects can enhance the effective cross section at the
freeze out temperature, so that, if the latter is normalized to that
of a thermal relic, the cross section in the dark star can be
suppressed.  In particular, the s-channel annihilation at one loop
through the exchange of the second KK Higgs with mass $m_{h^{(2)}}$ is
discussed in \cite{Kakizaki:2005en}. The canonical value for a thermal
relic $\langle \sigma v \rangle_{\rm fo} = 3 \times 10^{-26} {\rm cm^3
  s^{-1}}$ is obtained for a KK photon with mass $m_{\rm KK} \simeq
800$ GeV if the relation $m_{h^{(2)}}=2 m_{\rm KK}$ holds up to 5\%.
This new enhancement changes the cross section only by 10 to 20\%.
Therefore, the largest possible difference between $\langle \sigma v
\rangle_{\rm fo}$ and $\langle \sigma v \rangle_{\rm ds}$ can be at
most 10 to 20\%.  Unlike the MSSM, there is no p-wave suppression for
the KK photon annihilation and the cross section at freeze out
temperature is large enough due to the t-channel exchange of KK
fermions.  The same is true for KK neutrinos through the t-channel
exchange of KK Z and KK W bosons. The KK Z and KK W bosons have
similar masses, and in this case $f_Q \ll 1$ is not possible.  Thus
$f_Q \langle \sigma v \rangle_{\rm ds}/m_{\rm KK} \ge 10^{-32} {\rm
  cm^3} s^{-1}/ {\rm GeV}$ is safely satisfied.  Coannihilation with
the KK gluon is a last interesting possibility
\cite{Burnell:2005hm}. If the KK gluon and the KK photon are
degenerate with an accuracy much less than 1\%, the correct relic
density is obtained for $m_{\rm KK} \simeq 5$ TeV.  By comparing it to
the case without coannihilation ($m_{\rm KK} = 700$ GeV), one can see
that coannihilation with the KK gluon can enhance the effective cross
section by a factor of 50. Even in the worst scenario in which
coannihilation with the KK gluon is effective at the freeze out
temperature but is absent in the dark star, $ f_Q \langle \sigma v
\rangle_{\rm ds}/m_{\rm KK} \ge 4 \times 10^{-32} {\rm cm^3} s^{-1}/
{\rm GeV}$ (assuming a conservative value $f_Q=1/3$) and the dark star
can form.

As a consequence, if $\langle \sigma v \rangle_{\rm fo} = 3 \times
10^{-26} {\rm cm^3 s^{-1}}$ is imposed in order to explain the
observed relic density, the condition for the dark star formation is
always satisfied. We can conclude that KK dark matter that explains
the observed relic density can always form dark stars.

\section{Leptophilic Models}
\label{sec:lepto}

Leptophilic DM models \cite{leptophilic} have recently become popular
in order to explain simultaneously the excess in PAMELA
positrons~\cite{PAMELA}, the excess in Fermi-LAT electrons~\cite{FERMI}, as
well as the excellent agreement between the observed antiproton
spectrum and the corresponding standard expectation \cite{pbar}.  In order
to explain the excesses, the mass of the DM is also constrained to be
larger than about 100 GeV. This constraint might be more stringent if
the electron FERMI-Lat data are taken into account, $m > 400$ GeV.

In
leptophilic models the DM particles generically annihilate
exclusively to charged leptons, either of only one type (electrons,
taus or muons) or democratically to all the three families. It is also
possible to consider decays to neutrinos, but for simplicity we will
not consider this case which would simply imply a straightforward
generalization (see Section 5 for annihilation into neutrinos only). In particular, in this section we discuss the
case of democratic annihilation to the three lepton families. In this
case, using PYTHIA~\cite{pythia} one gets $f_Q\simeq 0.56$ almost
constant in the range 200 GeV$\lsim m\lsim$ 2 TeV.

In order to explain the PAMELA and Fermi-LAT excesses, large
annihilation cross sections $10^{-25}\lsim \langle  \sigma
v \rangle_{\rm gal}/{\rm cm^3 s^{-1}}\lsim 10^{-23}$ are needed at the velocity
of DM particles in our Galaxy, $v_{\rm gal}\simeq$ 300 km/s. Assuming
$\langle \sigma v \rangle_{\rm gal}=\langle \sigma v \rangle_{\rm fo}$ ($s$--wave annihilation)
these values are up to two orders of magnitude larger than the
value $\langle \sigma v \rangle_{\rm fo} \simeq 3 \times 10^{-26}$
cm$^3$ s$^{-1}$ compatible with a standard thermal relic abundance in agreement
with observations. Clearly, the case of $p$--wave suppression of
$\langle \sigma v \rangle_{\rm gal}$ would be even worse. Several mechanisms have
been devised in order to explain this discrepancy, such as a
non-thermal production of the DM particles, a non--standard evolution
history of the Universe or an enhancement of the annihilation cross
section at low velocities (Sommerfeld effect). In this sense
leptophilic models represent another interesting possibility in
connection with the formation of a dark star: a very large
annihilation cross section throughout the history of the Universe
and/or at the low temperatures where dark stars are formed.

\begin{figure}
\begin{center}
\includegraphics[bb=51 230 505
  645,width=0.7\linewidth]{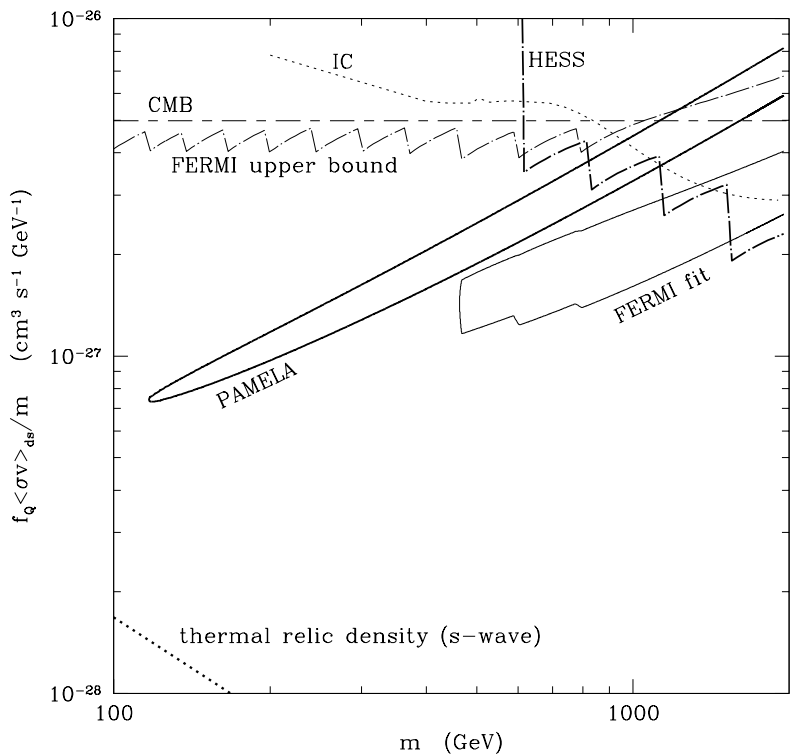}
\end{center}
\caption{ Combination $f_Q\langle\sigma v \rangle_{\rm ds}/m$ for a DM
  candidate annihilating democratically to charged leptons of the
  three families as a function of the mass $m$, assuming that
  $\langle\sigma v \rangle$ is independent of the temperature. The
  thick solid line contour shows the range of values compatible with
  the PAMELA positron excess~\cite{PAMELA}; the thin solid contour is
  the range compatible with the observed $e^{+}+e^{-}$ flux measured by
  FERMI--LAT~\cite{FERMI}; the thin solid open line is the $2\sigma$
  upper bound from the observed $e^{+}+e^{-}$ flux of FERMI; the thick
  solid open line is the $2\sigma$ upper bound from the $e^{+}+e^{-}$
  HESS measurement~\cite{HESS}; the dot--dashed line represents the
  upper bound on $\langle\sigma v\rangle$ from CMB~\cite{cmb}. The
  dotted line shows the upper bound on $\langle\sigma v \rangle$
  obtained by comparing the expected gamma--ray flux produced by
  Inverse Compton (IC) scattering of the final state leptons to the
  FERMI--LAT measurement of the diffuse gamma ray emission with
  subtraction of the expected standard
  background~\cite{FERMI_diffuse}. }
\label{fig:leptophilic_constraints}
\end{figure}

Assuming $\langle \sigma v \rangle_{\rm gal}$=$\langle \sigma v \rangle_{\rm ds}$, the previous
discussion implies that in a dark star:
\bea 7 \times 10^{-28}
{\rm cm^3 s^{-1} GeV^{-1}} \lsim f_Q \frac{\langle \sigma v \rangle_{\rm ds}}{m} \lsim 8 \times 10^{-27} {\rm cm^3 s^{-1} GeV^{-1}},
\label{eq:interval_leptophilic}
\eea
for $100 \ {\rm GeV} < m < 2$ TeV. This is shown in
Fig.~\ref{fig:leptophilic_constraints}, where along with the intervals
required to explain the PAMELA and Fermi/LAT excesses the present
constraints for the combination $f_Q\langle\sigma v \rangle_{\rm
  ds}/m$ are summarized as a function of $m$. In this plot we have
assumed that $\langle\sigma v \rangle$ does not depend on the
temperature, in order to directly compare constraints relative to
different epochs.

In particular, the thick and thin solid line closed contours show the
range of values compatible with the PAMELA positron
excess~\cite{PAMELA} and the FERMI--LAT $e^{+}+e^{-}$
data~\cite{FERMI}, respectively. On the other hand, the two solid open
lines represent conservative 2 $\sigma$ C.L.  upper bounds for
$f_Q\langle\sigma v \rangle_{\rm ds}/M_{\chi}$ obtained from the
flux of $e^{+}+e^{-}$ observed by FERMI \cite{FERMI} (thin line) and
the $e^{+}+e^{-}$ flux measured by HESS \cite{HESS} (thick line).

In the same Figure, we also plot with the dotted line the upper bounds
on $f_Q\langle\sigma v \rangle_{\rm ds}/m$ obtained by comparing the
expected gamma--ray flux produced by Inverse Compton (IC) scattering
of the final state leptons to the diffuse flux of gamma--rays measured
by FERMI at intermediate Galactic latitudes~\cite{FERMI_diffuse}.

Finally the long and short dashed line shows the upper bound on
$f_Q\langle\sigma v \rangle_{\rm ds}/m$ obtained by considering the
imprint on the Cosmic Microwave Background Radiation (CMB) from the
injection of charged leptons from DM annihilations at the
recombination epoch~\cite{cmb}.

It is clear from Fig.\ref{fig:leptophilic_constraints} that the
sizable values of the combination $f_Q\langle\sigma
v \rangle_{\rm ds}/M_{\chi}$ are compatible to the formation of a
dark star (Eq.\ref{eq:ds_condition}) when $\langle\sigma
v \rangle_{\rm ds}=\langle\sigma v \rangle_{\rm gal}$, with the
range $m\gsim$ 1 TeV disfavored~\cite{ic_scopel_chun} by several
constraints.  On the other hand, by assuming a Sommerfeld enhancement
of the annihilation cross section one may have $\langle\sigma
v \rangle_{\rm ds}\gsim\langle\sigma v \rangle_{\rm gal}$ depending
on whether the enhancement effect is already saturated at the velocity
$v\gsim 10$ km s$^{-1}$ inside the dark star, possibly implying in
this case an even more favorable situation for the dark star
formation. However, in presence of a non--saturated Sommerfeld
enhancement at the recombination epoch the CMB constraint could be
stronger, since at $z\simeq 1100$ DM particles are slower ($v\simeq
10^{-8} c\simeq 10^{-3}$ km s$^{-1}$~\cite{cmb}) than
inside a dark star. In this case leptophilic DM could explain the
PAMELA and Fermi/LAT excesses only by assuming a boost factor of
astrophysical origin such as clumpiness. In any case, barring some specific
cases such as the presence of resonances in the annihilation cross
section associated with bound states \cite{zavala} even in this
circumstance the bound in Eq.~(\ref{eq:ds_condition}) would be easily
verified and a dark star would be formed.

Thus we can also conclude that leptophilic models also satisfy the condition to form a dark star.

\section{Neutrinophilic Models}
\label{sec:neutro}

As discussed in the previous sections the most popular examples of
thermal dark matter candidates, namely the neutralino in the MSSM and
the KK photon or KK neutrino in Kaluza--Klein DM, can easily produce a
dark star, provided that the annihilation cross section at the
temperature of the dark star is similar to that at freeze out and is
not suppressed by mechanisms such as $p$--wave annihilation or by the
fact that the annihilation cross section is resonant at the freeze out
temperature but not inside the dark star. In this Section we wish to
generalize this statement to the general case of a thermal DM
candidate, discussing what are the minimal conditions to form a dark
star once $\svds$ is normalized to the
canonical value $\svfo=3\times 10^{-26}
{\rm cm^3/s}$. Moreover, we will also briefly comment on the case
$\svds<\svfo$.

For this purpose, we need to give a general discussion of the energy
fraction $f_Q$ released by DM annihilation into the gas, a quantity
that is in general model dependent.  Our approach to this problem is
to consider in this Section the most conservative case of a DM
candidate annihilating exclusively into neutrinos,
i.e. ``neutrinophilic'' Dark Matter.

Naively one would expect that the energy fraction of neutrinophilic DM
going into visible particles vanishes: DM annihilation would
generate only neutrinos that would escape from the collapsing gas
freely.  If this were indeed the case, neutrinophilic dark matter
annihilations would not be able to support a dark star phase.  However
Z and W bosons are expected to be produced from bremsstrahlung
radiation of the final state neutrinos, so some visible energy,
increasing with the mass of the DM particle, is expected to be
produced by the decay of the Z and/or W. Electroweak bremsstrahlung
in the annihilation of neutrinophilic DM has already been considered
in the context of DM indirect detection \cite{EWcorr}.

Let's first assume that neutrinophilic dark matter has an annihilation
cross section in the dark star equal to the cross section that
provides a thermal relic density, namely $\svds = 3 \times 10^{-26} {\rm cm^3 s^{-1}}$. At the
tree level the branching ratio to $\nu \bar{\nu}$ is $1$ if
bremsstrahlung radiation is neglected. However, when $2 m > m_W$
(or $m_Z$), on-shell production of W-bosons (or Z-bosons) dominates
the bremsstrahlung process, which can be viewed as a three body decay
followed by the subsequent decay of the W or Z gauge bosons. Since the
visible energy fraction of the W- and Z-boson decays is of order $1$,
one finds
\begin{equation}
 f_Q \sim \frac{g^2}{16\pi^2} \, \frac{E_W}{2m} > \frac{g^2}{16\pi^2} \,\frac{m_W}{2m}.
 \end{equation}
 Here $E_W$ is the energy of the W boson.
 In this case, as can be easily checked numerically, the condition for
 forming a dark star is always fulfilled. On the other hand, when
 $2m < m_W$, off-shell bremsstrahlung occurs, which for $2m
 \ll m_W$ can be treated as a four-body decay in the limit of a
 4-Fermi interaction.

\begin{figure}
\begin{center}
\includegraphics{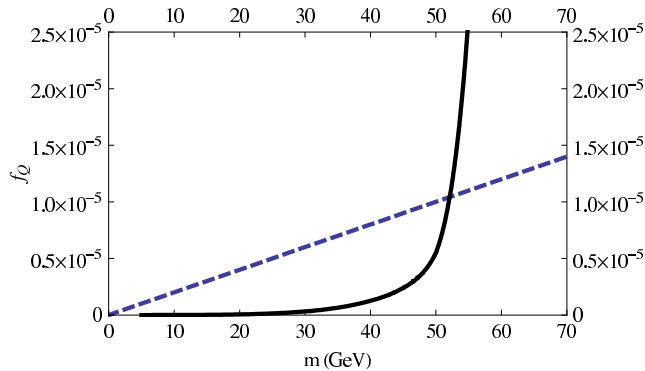}
\end{center}
\caption{Energy fraction $f_Q$ released by neutrinophilic dark matter
  annihilation inside a dark star as a function of the DM particle
  mass $m$. The black curve represents $f_Q$ for an ideal
  neutrinophilic DM model where the $W$ and $Z$ bremsstrahlung effects
  are calculated using PYTHIA~\cite{pythia}. The blue dashed line
  shows the constraint given in Eq.~(\protect\ref{eq:ds_condition}).
  Only models above the blue line can form a dark star.}
\label{fig:fv}
\end{figure}

 The visible energy fraction $f_Q$ is plotted as a function of the DM
 mass $m$ in Fig. \ref{fig:fv}. In this Figure we have
 used PYTHIA~\cite{pythia} to calculate the subsequent decay of the
 final state particles in the radiative correction, since the value of
 $f_Q$ critical for the formation of the dark star is near the
 threshold for the production of an on-shell W-boson, where a narrow
 width approximation or a 4-Fermi interaction are not reliable.  From
 Fig.\ref{fig:fv} we can conclude that a dark star can be formed if
 the mass of the neutrinophilic dark matter is larger than $\sim50$
 GeV.

 If the DM particle is a scalar or a Majorana fermion, it is possible
 that the annihilation cross section in the dark star is significantly
 different from the cross section at decoupling. A simple example is a
 scalar neutrinophilic dark matter particle $\phi$ annihilating by
 $t$/$u$-channels through the exchange of a heavy fermion. In
 particular, the $s$-wave contribution vanishes if the heavy fermion
 mediating the annihilation is a Dirac fermion and if it interacts
 chirally with the dark matter $\phi$ and the neutrinos, i.e.
\begin{equation}
{\cal L}\sim g \phi \Psi_L \nu  +M\Psi_L\Psi_R+{\rm h.c.}+m^2|\phi|^2,
\end{equation}
where $M$ is the mass of the heavy fermion and $m$ is the mass of the
dark matter scalar. In this case, the annihilation cross section is
purely $p$-wave and is given by
\begin{equation}
\sigma v=v^2 \frac{|g|^4}{16\pi}\, \frac{m^2( (M^2+m^2)^2 + 2m^2M^2)}{(m^2+M^2)^4} .
\end{equation}
In this case, since the average velocity of the dark matter in the
dark star is $\sim 30$ km/s, one has $\langle \sigma v\rangle_{\rm
  ds}=\langle \sigma v\rangle_{\rm fo}(v_{\rm ds}/v_{\rm fo})^2
\sim\langle \sigma v\rangle_{\rm fo}v_{\rm ds}^2(m/T_{\rm
  fo})\sim10^{-32}{\rm cm^3/s}$. So $p$-wave annihilating
neutrinophilic DM cannot make a dark star.

Notice that in our evaluation of neutrinophilic DM annihilation we
have not included internal bremsstrahlung of charged
particles. Internal bremsstrahlung would increase the annihilation
cross section of neutrinophilic dark matter, and extend the mass limit
for the formation of dark stars to values below $\sim50$ GeV, but
would depend on the specific particle physics model.

Strictly speaking a purely neutrinophilic DM model is not natural, since
in most specific scenarios other tree--level decay channels
contributing to the visible energy are usually expected, and a
contribution of the latter at the level of $\sim10^{-5}$ level or
larger would be sufficient to form a dark star. Thus neutrinophilic
DM represents a limiting case, allowing to show that, as long as
$\svds=\svfo$, any thermal DM candidate heavier
than $m\gsim 50$ GeV can lead to the formation of a dark star.

\section{Conclusions}

The first stars to form in the Universe may be powered by the
annihilation of weakly interacting dark matter
particles~\cite{dark_stars_prl}.  In this paper we explored several
popular examples of thermal dark matter models in order to discuss
whether they can satisfy the conditions for the formation of a dark
star: the neutralino in an effective MSSM scenario; leptophilic models
that might explain recent observations in cosmic rays; the KK-photon
and the KK-neutrino in UED models; a conservative neutrinophilic model
where the dark matter particles annihilate exclusively to neutrinos.
We find that in general models with thermal dark matter lead to the
formation of dark stars, with few notable exceptions: heavy
neutralinos in the presence of coannihilations; annihilations that are
resonant at dark matter freeze-out but not in dark stars;
neutrinophilic dark matter lighter than about 50 GeV. In particular
the discussion of the latter conservative scenario allows us to
conclude that a thermal DM candidate in standard Cosmology always
forms a dark star as long as its mass is heavier than $\simeq$ 50 GeV
and the thermal average of its annihilation cross section is the same
at the decoupling temperature and during the dark star formation, as
for instance in the case of a cross section with a non--vanishing
$s$-wave contribution.

Therefore, we can conclude that the formation of a first generation of
stars powered by dark matter annihilation is an almost inevitable
consequence of thermal dark matter when a standard thermal history of
the Universe is assumed and if the mechanism of
Ref.~\cite{dark_stars_prl} is at work. So a dark star is always there
whenever there is thermal dark matter.

\section{Acknowledgments}
This work is supported by KRF-2008-313-C00162 (HDK), by NRF with CQUEST
grant 2005-0049049 (HDK, SS), and partially by NSF award PHY-0456825 and NASA grant NNX09AT70G (PG).


\begin{thebibliography}{999}

\bibitem{dark_stars_prl}
  D.~Spolyar, K.~Freese and P.~Gondolo,
  Phys.\ Rev.\ Lett.\  {\bf 100}, 051101 (2008)
  [arXiv:0705.0521 [astro-ph]].

\bibitem{Freese:2008hb}
  K.~Freese, P.~Gondolo, J.~A.~Sellwood and D.~Spolyar,
  Astrophys.\ J.\  {\bf 693} (2009) 1563
  [arXiv:0805.3540 [astro-ph]].
  
 \bibitem{Spolyar:2009nt}
  D.~Spolyar, P.~Bodenheimer, K.~Freese and P.~Gondolo,
  Astrophys.\ J.\  {\bf 705} (2009) 1031
  [arXiv:0903.3070 [astro-ph.CO]].

\bibitem{Natarajan:2008db}
  A.~Natarajan, J.~C.~Tan and B.~W.~O'Shea,
  Astrophys.\ J.\  {\bf 692}, 574 (2009)
  [arXiv:0807.3769 [astro-ph]].


\bibitem{yoshida}
  N.~Yoshida, K.~Omukai, L.~Hernquist and T.~Abel,
  Astrophys.\ J.\  {\bf 652}, 6 (2006)
  [arXiv:astro-ph/0606106].


\bibitem{Griest:1990kh}
  K.~Griest and D.~Seckel,
  Phys.\ Rev.\  D {\bf 43}, 3191 (1991).


\bibitem{Servant:2002aq}
  G.~Servant and T.~M.~P.~Tait,
  Nucl.\ Phys.\  B {\bf 650}, 391 (2003)
  [arXiv:hep-ph/0206071].


\bibitem{Kakizaki:2005en}
  M.~Kakizaki, S.~Matsumoto, Y.~Sato and M.~Senami,
  Phys.\ Rev.\  D {\bf 71}, 123522 (2005)
  [arXiv:hep-ph/0502059].
  

\bibitem{Burnell:2005hm}
  F.~Burnell and G.~D.~Kribs,
  Phys.\ Rev.\  D {\bf 73}, 015001 (2006)
  [arXiv:hep-ph/0509118].
  K.~Kong and K.~T.~Matchev,
  JHEP {\bf 0601}, 038 (2006)
  [arXiv:hep-ph/0509119].

\bibitem{leptophilic}
See for instance:
 P.~J.~Fox and E.~Poppitz,
  Phys.\ Rev.\  D {\bf 79}, 083528 (2009)
  [arXiv:0811.0399 [hep-ph]];
 R.~Harnik and G.~D.~Kribs,
  Phys.\ Rev.\  D {\bf 79}, 095007 (2009)
  [arXiv:0810.5557 [hep-ph]];
 N.~Arkani-Hamed, D.~P.~Finkbeiner, T.~R.~Slatyer and N.~Weiner,
  Phys.\ Rev.\  D {\bf 79}, 015014 (2009)
  [arXiv:0810.0713 [hep-ph]].

\bibitem{PAMELA}
  O.~Adriani {\it et al.}  [PAMELA Collaboration],
  Nature {\bf 458}, 607 (2009)
  [arXiv:0810.4995 [astro-ph]].

\bibitem{FERMI}
  A.~A.~Abdo {\it et al.}  [The Fermi LAT Collaboration],
  Phys.\ Rev.\ Lett.\  {\bf 102}, 181101 (2009)
  [arXiv:0905.0025 [astro-ph.HE]].


\bibitem{pbar}
  O.~Adriani {\it et al.}  [PAMELA Collaboration],
  Phys.\ Rev.\ Lett.\  {\bf 102}, 051101 (2009)
  [arXiv:0810.4994 [astro-ph]].

\bibitem{ic_scopel_chun}
  E.~J.~Chun, J.~C.~Park and S.~Scopel,
  JCAP {\bf 1002}, 015 (2010)
  [arXiv:0911.5273 [hep-ph]].


\bibitem{HESS}
  F.~Aharonian {\it et al.}  [H.E.S.S. Collaboration],
  Phys.\ Rev.\ Lett.\  {\bf 101}, 261104 (2008)
  [arXiv:0811.3894 [astro-ph]];
  arXiv:0905.0105 [astro-ph.HE].


\bibitem{cmb}
 S.~Galli, F.~Iocco, G.~Bertone and A.~Melchiorri,
  Phys.\ Rev.\  D {\bf 80}, 023505 (2009)
  [arXiv:0905.0003 [astro-ph.CO]];
  T.~R.~Slatyer, N.~Padmanabhan and D.~P.~Finkbeiner,
  Phys.\ Rev.\  D {\bf 80}, 043526 (2009)
  [arXiv:0906.1197 [astro-ph.CO]];
 N.~Padmanabhan and D.~P.~Finkbeiner,
  Phys.\ Rev.\  D {\bf 72}, 023508 (2005)
  [arXiv:astro-ph/0503486].


\bibitem{FERMI_diffuse}
 T.~A.~Porter for the Fermi Collaboration,
  arXiv:0907.0294 [astro-ph.HE].


\bibitem{zavala}
  J.~Zavala, M.~Vogelsberger and S.~D.~M.~White,
  arXiv:0910.5221 [astro-ph.CO].


\bibitem{EWcorr}
  N.~F.~Bell, J.~B.~Dent, T.~D.~Jacques and T.~J.~Weiler,
  Phys.\ Rev.\  D {\bf 78} (2008) 083540
  [arXiv:0805.3423 [hep-ph]].

\bibitem{pythia}
  T.~Sjostrand, S.~Mrenna and P.~Z.~Skands,
  JHEP {\bf 0605} (2006) 026
  [arXiv:hep-ph/0603175].


\bibitem{effMSSM}
  A.~Bottino, F.~Donato, N.~Fornengo and S.~Scopel,
  Phys.\ Rev.\  D {\bf 63} (2001) 125003
  [arXiv:hep-ph/0010203]


\end{thebibliography}
\end{document}